\documentclass{article}
\usepackage{arxiv}

\usepackage{graphicx}
\usepackage{mathrsfs}

\title{\boldmath Application of Fermion Quantum Number $F$ and Unit Electroweak Charge $F_0$ in the Electroweak Theory}

\author{
 Xin-Hua Ma \\
  Key Laboratory of Particle Astrophyics, Institute of High Energy Physics, Chinese Academy of Sciences, \\ 100049 Beijing, China\\
  TIANFU Cosmic Ray Research Center, Chengdu, Sichuan, China \\
  \texttt{maxh@ihep.ac.cn} \\
}
\begin{document}
\maketitle
\begin{abstract}
In the previous work, the traditional flavor-related additive quantum numbers were substituted by the fermion quantum number and the unit electroweak charge, which are, same as the electric charge, conserved in electromagnetic interaction, weak interaction and strong interaction. The empirical selection rules in electroweak interaction were found that they can be interpreted by conservation of the fermion quantum number. In this paper, the weak isospin and the weak hypercharge in the electroweak theory are demonstrated to be equivalent to half of the fermion quantum number and the unit electroweak charge respectively. Therefore, we can unify the traditional flavor-related additive quantum numbers and the additive quantum numbers in the electroweak theory by using only two numbers. Furthermore, we can take a new point of view on the electroweak theory by using these two numbers.
\\[0.2cm]
\end{abstract}

\section{Introduction}\label{sec:1}

\begin{table*}[b!]
\centering
\begin{tabular}{c|ccllllccccclc}
\hline
   fermion       & $Q$  & $B$ &$L_e$ &$L_\mu$ &$L_\tau$ &$L$& $I$ & $I_3$ & $S$&$C$& $B^*$ &$T$& $Y$  \\
\hline
	 $d$        & -1/3 & 1/3 & 0    & 0      & 0       & 0 & 1/2 & -1/2  & 0  & 0 & 0     & 0 &  1/3 \\
	 $u$        &  2/3 & 1/3 & 0    & 0      & 0       & 0 & 1/2 &  1/2  & 0  & 0 & 0     & 0 &  1/3 \\
	 $s$        & -1/3 & 1/3 & 0    & 0      & 0       & 0 & 0   & 0     & -1 & 0 & 0     & 0 & -2/3 \\
	 $c$        &  2/3 & 1/3 & 0    & 0      & 0       & 0 & 0   & 0     & 0  & 1 & 0     & 0 &  4/3 \\
	 $b$        & -1/3 & 1/3 & 0    & 0      & 0       & 0 & 0   & 0     & 0  & 0 & -1    & 0 & -2/3 \\
	 $t$        &  2/3 & 1/3 & 0    & 0      & 0       & 0 & 0   & 0     & 0  & 0 & 0     & 1 &  4/3 \\
      $e$        & -1   & 0   & 1    & 0      & 0       & 1 & 0   & 0     & 0  & 0 & 0     & 0 & 0    \\
	 $\nu_e$    & 0    & 0   & 1    & 0      & 0       & 1 & 0   & 0     & 0  & 0 & 0     & 0 & 0    \\
      $\mu$      & -1   & 0   & 0    & 1      & 0       & 1 & 0   & 0     & 0  & 0 & 0     & 0 & 0    \\
	 $\nu_\mu$  & 0    & 0   & 0    & 1      & 0       & 1 & 0   & 0     & 0  & 0 & 0     & 0 & 0    \\
	 $\tau$     & -1   & 0   & 0    & 0      & 1       & 1 & 0   & 0     & 0  & 0 & 0     & 0 & 0    \\
	 $\nu_\tau$ & 0    & 0   & 0    & 0      & 1       & 1 & 0   & 0     & 0  & 0 & 0     & 0 & 0    \\
\hline
\end{tabular}
\caption{Values of the traditional flavor-related additive quantum numbers and the additive quantum numbers conserved in all three interactions. The antifermions have the additive quantum numbers with the same absolute values as the fermions but the opposite sign of the fermions, except that the antiquarks have same $I$ as the quarks.\label{traditional}}
\end{table*}

At the most elementary particle level, the fermions include the quarks and the leptons. During development of elementary particle physics, it was found that the fermions have several additive quantum numbers defined as the electric charge $Q$ (with unit of the electron charge $e$), the baryon quantum number $B$, the lepton quantum number $L$, the hypercharge $Y$, the isospin $I$ with the isospin projection $I_3$ and four flavor quantum numbers including $S$, $C$, $B^*$ and $T$. Only $Q$, $B$ and $L$ are conserved in electromagnetic interaction, strong interaction and weak interaction. Especially, weak interaction changes the lepton only into itself or into the other member of the same generation in the lepton family, so the leptons have three individually conserved lepton quantum numbers $L_e$, $L_\mu$ and $L_\tau$. $I$ is conserved in strong interaction, but not conserved in either electromagnetic interaction or weak interaction. $Y$, $I_3$, $S$, $C$, $B^*$ and $T$ are conserved in electromagnetic interaction and strong interaction, but not conserved in weak interaction. In weak interaction, the systematic way in which the symmetry is broken has led to several empirical selection rules. On the other hand, independently from the traditional flavor-related additive quantum numbers (including $Y$, $I$, $I_3$, $S$, $C$, $B^*$ and $T$), the weak isospin $I_w$ and its projection $I_{w3}$ and the weak hypercharge $Y_w$ were inducted in the electroweak theory, also known as the Glashow-Weinberg-Salam (GWS) theory. The antifermions have the additive quantum numbers with the same absolute values as the fermions but the opposite sign of the fermion, except that the antiquarks have same $I$ as the quarks. Literature about the elementary particle physics can be found in, e.g., ~\cite{Zhang1}\cite{Zhang2}\cite{Kim}\cite{Nagashima1}\cite{Nagashima2} .

In fact, there exist several problems inside the concepts of the traditional additive quantum numbers. $I$ and $I_3$, having absolute value 1/2, are the traditional flavor-related quantum numbers of u and d quark essentially, but different from the flavor quantum numbers $S$, $C$, $B^*$ and $T$ for the other quarks, having absolute value 1. Why are the u and d quark so `unique' in the quark family? Moreover, $Y$ is sum of $B$ and the flavor quantum numbers which is a strange mixture between $B$ conserved in all three interactions and four flavor quantum numbers conserved in only part of three interactions. And it was unknown why the empirical selection rules in weak interaction can work. For electroweak interaction $I_{w3}$ and $Y_w$ were analogical to $I$ and $Y$ by using the Gell-Mann-Nishijima formula, but values of $I_{w3}$ and $Y_w$ are completely different with $I$ and $Y$ respectively except for the left-handed u and d quark. It was not considered before if there is a relation between the traditional flavor-related additive quantum numbers and the additive quantum numbers in electroweak theory ($I_{w3}$ and $Y_w$) or not.

Recently, the fermion quantum number $F$ was brought forward to replace $I$, $I_3$, $S$, $C$, $B^*$ and $T$ so that $u$ and $d$ quark become `common' as the other four quarks, and $F_0$ (combination of $B$ and $L$) was brought forward to replace $Y$ \cite{Ma1}. Same as $Q$, both $F$ and $F_0$ are conserved in electromagnetic interaction, strong interaction and weak interaction. The formula of the relation between $Q$, $F$ and $F_0$ for all fermions is more general than the Gell-Mann-Nishijima formula \cite{Ma1}. Furthermore, it was found that all the miscellaneous selection rules in all types of weak interaction can be interpreted by conservation of $F$, and even why some selections have not been observed in experiments can be explained by not conservation of $F$ in the cases \cite{Ma2}. In this paper, I apply $F$ and $F_0$ in the electroweak theory and deduce Lagrangian of the electroweak interaction in new type.

\section{Traditional Flavor-related Additive Quantum Numbers and Additive Quantum Numbers in the Electroweak Theory }

Values of the traditional flavor-related quantum numbers, including $Y$, $I$, $I_3$, $S$, $C$, $B^*$ and $T$, and the additive quantum numbers conserved in all three interactions, including $Q$, $B$, $L_e$, $L_\mu$, $L_\tau$ and $L$, are listed in Table \ref{traditional}. The relation between $Q$, $I_3$ and $Y$ was given by the Gell-Mann-Nishijima formula \cite{GN1}\cite{GN2}\cite{GN3} (in extended form):

\begin{equation}
Q=I_3+Y/2
\label{eqnGN}
\end{equation}
where $Y = B + S + C + B^{*} + T$.

The status that different traditional flavor-related additive quantum numbers were given to different quarks is due to historical development of discovery of six quarks: first of all were $u$ and $d$ quark for which the isospin projection $I_3$ were given as a pair, then $S$ for $s$ quark, and then $C$, $B^*$ and $T$ for $c$, $b$ and $t$ quark successively. $Y$ was extended passively. Although the numbers were launched one by one following the experimental results, their quantities have revealed their natural connotations: absolute values of $I_3$, $S$, $C$, $B^*$ and $T$ are only 1/2 or 1 and signs are alternative between plus and minus.

Moreover, it has been misunderstood for a long time that the traditional flavor-related additive quantum numbers are the numbers “only” for strong interaction. It is not true, because simultaneously they are used to describe phenomena in weak interaction, but unfortunately, all of them are not conserved in weak interaction. As a compromise, weak interaction had to be separated into three types: pure leptonic, pure hadronic (or nonleptonic) or fermionic (or semi-leptonic, semi-hadronic), and the traditional flavor-related additive quantum numbers had to be organized into a series of complex selection rules.

On the other hand, the additive quantum numbers in the electroweak theory ($Y_w$ and $I_{w3}$) were launched to describe about electroweak interaction, and they are obviously also related to different flavors, but they were not connected with the traditional flavor-related additive quantum numbers.

In the GWS theory, weak interaction in its original form, that is, before mixing and spontaneous symmetry breakdown, has chiral $SU(2)$ symmetry. The weak interaction and electromagnetic interaction are based on a mixture of $SU(2)$ and $U(1)$. All the left-handed fermions constitute doublets. All the right-handed fermions belong to $SU(2)$ singlets. In this section, the form of the variables and the equations refers to \cite{Nagashima2}, except that lower corner mark `w' for `weak' is added explicitly. All the leptons can be classified by their weak isospin component as
\begin{equation}
\begin{array} {l}
  \left\{ \begin{array} {l} I_{w3}=1/2 \\ I_{w3}=-1/2  \end{array} \right. \\
  I_{w3}=0
\end{array}
\quad
\begin{array} {l}
   \Psi_L= \left( \begin{array} {l} \nu_e \\ e \end{array} \right)_L ,
     \left( \begin{array} {l} \nu_{\mu} \\ \mu \end{array} \right)_L ,
     \left( \begin{array} {l} \nu_{\tau} \\ \tau \end{array} \right)_L . \\
    e_R, \mu_R, \tau_R
\end{array}
\label{psilepton}
\end{equation}
It is an experimental result: only the left-handed neutrinos exist, but the right-handed neutrinos are not detected. Non-existence of the right-handed neutrinos is regarded as the reason of parity violation in weak interaction including the neutrinos. For the quarks
\begin{equation}
\begin{array} {l}
  \left\{ \begin{array} {l} I_{w3}=1/2 \\ I_{w3}=-1/2  \end{array} \right. \\
  I_{w3}=0
\end{array}
\quad
\begin{array} {l}
   \Psi_L= \left( \begin{array} {l} u \\ d' \end{array} \right)_L ,
     \left( \begin{array} {l} c \\ s' \end{array} \right)_L ,
     \left( \begin{array} {l} t \\ b' \end{array} \right)_L\\
    u_R, d_R, c_R, s_R, t_R, b_R
\end{array}
\label{Psiquark}
\end{equation}
where ($d'$,$s'$,$b'$) are Cabibbo-Kobayashi-Maskawa (CKM) rotated fields. Relation between $Q$, $I_{w3}$ and $Y_w$ is given by the Gell-Mann-Nishijima-like formula
\begin{equation}
Q=I_{w3}+Y_w/2 .
\label{eqnGNw}
\end{equation}
Values of $I_{w3}$ and $Y_w$ of the fermions are explicitly listed in Table \ref{GWS-left} and \ref{GWS-right}. The right-handed neutrinos do not exist so that the values for the right-handed neutrinos are zero. Comparing Table \ref{GWS-left}, \ref{GWS-right} and \ref{traditional}, we can see that $I_{w3}$ and $Y_w$ are completely different from $I_3$ and $Y$ except for $d'_L$ and $u_L$.

\begin{table}[htbp]
\centering
\begin{tabular}{c|cccccccccccc}
\hline
fermion & $d'_L$ & $u_L$  & $s'_L$ & $c_L$ & $b'_L$ & $t_L$ & $e_L$ & $\nu_{e L}$ & $\mu_L$ & $\nu_{\mu L}$ &  $\tau_L$ & $\nu_{\tau L}$   \\
\hline
$Q$ & -1/3 & 2/3 & -1/3 & 2/3 & -1/3 & 2/3 & -1 & 0 & -1 & 0 & -1 & 0 \\
$I_{w3}$ & -1/2 & 1/2 & -1/2 & 1/2 & -1/2 & 1/2 & -1/2 & 1/2 & -1/2 & 1/2 & -1/2 & 1/2 \\
$Y_{w}$ & 1/3 & 1/3 & 1/3 & 1/3 & 1/3 & 1/3 & -1 & -1 & -1 & -1 & -1 & -1 \\
\hline
\end{tabular}
\caption{$I_{w3}$ and $Y_w$ of the left-handed fermions in the GWS theory. The antifermions have the additive quantum numbers with the same absolute values as the fermions but the opposite sign of the fermions.\label{GWS-left}}
\end{table}

\begin{table}[htbp]
\centering
\begin{tabular}{c|cccccccccccc}
\hline
fermion  & $d_R$ & $u_R$  & $s_R$ & $c_R$ & $b_R$ & $t_R$ & $e_R$ & $\nu_{e R}$ & $\mu_R$ & $\nu_{\mu R}$ &  $\tau_R$ & $\nu_{\tau R}$     \\
\hline
$Q$ & -1/3 & 2/3 & -1/3 & 2/3 & -1/3 & 2/3 & -1 & 0 & -1 & 0 & -1 & 0 \\
$I_{w3}$ & 0 & 0 & 0 & 0 & 0 & 0 & 0 & 0 & 0 & 0 & 0 & 0 \\
$Y_{w}$ & -2/3 & 4/3 & -2/3 & 4/3 & -2/3 & 4/3 & -2 & 0 & -2 & 0 & -2 & 0 \\
\hline
\end{tabular}
\caption{$I_{w3}$ and $Y_w$ of the right-handed fermions in the GWS theory. The antifermions have the additive quantum numbers with the same absolute values as the fermions but the opposite sign of the fermions.\label{GWS-right}}
\end{table}

In the original Lagrangian of the electroweak interaction, the part of interaction of the gauge boson with the fermion
\begin{equation}
-\mathcal{L}_{Wff}=\rm i\bar{\Psi} \gamma^\mu D_\mu\Psi
\label{Lwff1}
\end{equation}
is contained in the covariant derivative
\begin{equation}
D_\mu=\partial_\mu+ \rm i g_W\mathbf{W}_\mu\cdot\mathbf{t}+ \rm i\frac{g_B}{2}Y_wB_\mu
\label{Dmu}
\end{equation}
where $\gamma^\mu$ ($\mu$ = 0,1,2,3) are Dirac $\gamma$ matrices, $\mathbf{W}_\mu$ is the $SU(2)$ gauge field, $B_{\mu}$ is the gauge boson of $U(1)$, $g_W$ and  $g_B$ are coupling strength of $\mathbf{W}_\mu$ and $B_{\mu}$ respectively, $\mathbf{t}=\tau/2$ is weak isospin operator, and $\tau$ is the Pauli $2 \times 2$ matrix. By using the Weinberg angle $\theta_W$, $g_W$ and $g_B$ has relation in
\begin{equation}
   \rm tan \theta_W=\frac{g_B}{g_W} \cdot
\label{thetawgBgw}
\end{equation}

After mixing $SU(2)$ and $U(1)$, it can be deduced
\begin{equation}
-\mathcal{L}_{Wff}=\mathcal{L}_W+\mathcal{L}_Z+\mathcal{L}_{em}
\label{LwffLLL}
\end{equation}
which is composed of the charged current weak interaction part
\begin{equation}
   \mathcal{L}_W=\frac{g_W}{\sqrt{2}}\bar{\Psi}_L\gamma^\mu(W^+_\mu\tau_++W^-_\mu\tau_-)\Psi_L
\label{LwI3Y}
\end{equation}
the neutral current weak interaction part
\begin{equation}
   \mathcal{L}_Z=g_Z\bar{\Psi}\gamma^\mu(I_{w3}-Qs^2_W)\Psi Z_\mu
\label{LZI3Y}
\end{equation}
and the electromagnetic interaction part
\begin{equation}
   \mathcal{L}_{em}=e\bar{\Psi}\gamma^\mu Q\Psi A_\mu
\label{LemI3Y}
\end{equation}
where $W_\mu^+$ and $W_\mu^-$ are charged boson field operators, $Z_\mu$ is neutral boson field operator, $A_\mu$ is electromagnetic field operator, $\tau_\pm=(\tau_1 \pm {\rm i}\tau_2)/2$, $s_W=sin\theta_W$, $c_W=cos\theta_W$, and $e$ and $g_Z$ are coupling constant of $A_\mu$ and $Z_\mu$ respectively defined in terms of $g_W$ and  $\theta_W$ as
\begin{equation}
\begin{array} {l}
e=g_W s_W , \qquad  g_Z=g_W/c_W.
\end{array}
\label{egZ}
\end{equation}
Coupling types which appear in the Feynman amplitude rule are given by the matrix element ${\rm i}\mathcal{L}_{Wff}$, so, omitting the field operators and attaching a suffix to differentiate the fermion flavor, interaction vertices are expressed as
\begin{equation}
\gamma - ff: -{\rm i} e\gamma^\mu Q_f
\label{gff}
\end{equation}
\begin{equation}
W^\pm - ff: -{\rm i}\frac{g_W}{2\sqrt{2}}\gamma^\mu (1-\gamma^5)
\label{Wff}
\end{equation}
\begin{equation}
Z - ff: -{\rm i}\frac{g_Z}{2}\gamma^\mu (v_f-a_f\gamma^5)
\label{ZffIY}
\end{equation}
where $\gamma^5$ is chirality operator, $f$ = $e, \mu, \tau$ or $d, u, s, c, b, t$, and
\begin{equation}
\begin{array} {l}
    v_f=I_{w3f}-2Q_fs_W^2 , \qquad   a_f=I_{w3f} .
\end{array}
\label{vfafI3Y}
\end{equation}

\section{Application of $F$ and $F_0$ in the Electroweak Theory}

Not like the traditional flavor-related traditional additive quantum numbers, $F$ for all fermions has simple value \cite{Ma1}:
\begin{equation}
F= \left\{ \begin{array} {r@{\quad:\quad}l} -1 & \mbox{fermion is 'd' type} \\ +1  & \mbox{fermion is 'u' type.} \end{array} \right.
\label{eqnFHHL}
\end{equation}
The relation between electric charge $Q$,  $F_0$ and $F$ is simple and direct  \cite{Ma1}
\begin{equation}
Q=\frac{F}{2}+\frac{F_0}{2}
\label{eqnQF}
\end{equation}
where F$_0$ is
\begin{equation}
 F_0= \left\{ \begin{array} {r@{\quad:\quad}l} B = +1/3 & \mbox{fermion is quark} \\ -L = -1  & \mbox{fermion is lepton.} \end{array} \right.
\label{eqnDFv}
\end{equation}

$F$ and $F_0$, same as $Q$, are conserved in electromagnetic interaction, strong interaction and weak interaction. The Formula (Equation \ref{eqnQF}) of the relation between $Q$, $F$ and $F_0$ is for all fermions, more general than the Gell-Mann-Nishijima formula (Equation \ref{eqnGN}), and only adopts the additive quantum numbers conserved in all three interactions. Details are described in the previous work \cite{Ma1}. Values of $Q$, $F$ and $F_0$ are listed in Table \ref{F-F0}.

Description of weak interaction becomes simpler after $F$ is inducted: It is proved that conservation of $F$ determines the selection rules in all types of weak interaction, including pure leptonic interaction, pure hadronic interaction and fermionic weak interaction  \cite{Ma2}. As an example to show how conservation of $F$ works, in $\beta$ decay
\begin{equation}
n \rightarrow p+ e^- + \bar{\nu_e}
\label{eqnbeta}
\end{equation}
or in the quark terms
\begin{equation}
ddu \rightarrow uud + e^- + \bar{\nu_e}
\label{eqnbeta2}
\end{equation}
or in the simplified term
\begin{equation}
d \rightarrow u + e^- + \bar{\nu_e}
\label{eqnbeta3}
\end{equation}
because $F$ of $d$ quark, $u$ quark, $e^-$ and $\bar{\nu_e}$ is -1, +1, -1 and -1 respectively (Table \ref{F-F0}), $F$ changes from -1 into +1-1-1 = -1, so $\Delta F = 0$, i.e., $F$ is conserved.

Compared with miscellaneous selection rules, conservation of F is more distinct and rather simpler for judgment on how the fermions react in all types of weak interaction. Moreover, even the reason why some selections have not been observed in experiments (e.g., $\Delta S = -\Delta Q$ in fermionic weak interaction) can be explained by not conservation of $F$ in the cases. Details are described in the previous work \cite{Ma2}. Conservation of $F$ and its function in the selection rules of weak interaction indicate that both the quarks and the leptons must be considered together in weak interaction, but not separately, and after it, interpretation of behaviour of the fermions  becomes simplified.

\begin{table}[htbp]
\centering
\begin{tabular}{c|cccccccccccc}
\hline
fermion & $d$ & $u$  & $s$ & $c$ & $b$ & $t$ & $e$ & $\nu_{e}$ & $\mu$ & $\nu_{\mu}$ &  $\tau$ & $\nu_{\tau}$   \\
\hline
$Q$ & -1/3 & 2/3 & -1/3 & 2/3 & -1/3 & 2/3 & -1 & 0 & -1 & 0 & -1 & 0 \\
$F$ & -1 & 1 & -1 & 1 & -1 & 1 & -1 & 1 & -1 & 1 & -1 & 1 \\
$F_{0}$ & 1/3 & 1/3 & 1/3 & 1/3 & 1/3 & 1/3 & -1 & -1 & -1 & -1 & -1 & -1 \\
\hline
\end{tabular}
\caption{Values of $Q$, $F$ and $F_0$. The antifermions have the additive quantum numbers with the same absolute values as the fermions but with opposite sign of the fermions.\label{F-F0}}
\end{table}

By comparing Table \ref{F-F0} with Table \ref{GWS-left}, it is demonstrated that for the left-handed fermions,
\begin{equation}
   I_{w3}=\frac{F}{2}
\label{Iw3F}
\end{equation}
and
\begin{equation}
   Y_{w}=F_0 .
\label{YwF0}
\end{equation}
That is, the weak isospin projection $I_{w3}$ and the weak hypercharge $Y_w$ in the electroweak theory are equivalent to $F/2$ and $F_0$ respectively.  By extending $F$ and $F_0$ to the right-handed fermions with Equation \ref{eqnQF} and $F=0$, we obtain values of $F$ and $F_0$ of all fermions listed in Table \ref{Ma-left-right}.

\begin{table}[htbp]
\centering
\begin{tabular}{c|cccccccccccc}
\hline
$f_L$ & $d'_L$ & $u_L$  & $s'_L$ & $c_L$ & $b'_L$ & $t_L$ & $e_L$ & $\nu_{e L}$ & $\mu_L$ & $\nu_{\mu L}$ &  $\tau_L$ & $\nu_{\tau L}$   \\
\hline
$Q$ & -1/3 & 2/3 & -1/3 & 2/3 & -1/3 & 2/3 & -1 & 0 & -1 & 0 & -1 & 0 \\
$F$ & -1 & 1 & -1 & 1 & -1 & 1 & -1 & 1 & -1 & 1 & -1 & 1 \\
$F_{0}$ & 1/3 & 1/3 & 1/3 & 1/3 & 1/3 & 1/3 & -1 & -1 & -1 & -1 & -1 & -1 \\
\hline
\end{tabular}
\begin{tabular}{c|cccccccccccc}
\hline
$f_R$  & $d_R$ & $u_R$  & $s_R$ & $c_R$ & $b_R$ & $t_R$ & $e_R$ & $\nu_{e R}$ & $\mu_R$ & $\nu_{\mu R}$ &  $\tau_R$ & $\nu_{\tau R}$     \\
\hline
$Q$ & -1/3 & 2/3 & -1/3 & 2/3 & -1/3 & 2/3 & -1 & 0 & -1 & 0 & -1 & 0 \\
$F$ & 0 & 0 & 0 & 0 & 0 & 0 & 0 & 0 & 0 & 0 & 0 & 0 \\
$F_{0}$ & -2/3 & 4/3 & -2/3 & 4/3 & -2/3 & 4/3 & -2 & 0 & -2 & 0 & -2 & 0 \\
\hline
\end{tabular}

\caption{$Q$, $F$ and $F_0$ extended to the left-handed and right-handed fermions in the electroweak theory. The antifermions have the additive quantum numbers with the same absolute values as the fermions but the opposite sign of the fermions.\label{Ma-left-right}}
\end{table}

In term of F, the leptons can be classified as
\begin{equation}
\begin{array} {l}
  \left\{ \begin{array} {l} F=1 \\ F=-1  \end{array} \right. \\
  F=0
\end{array}
\quad
\begin{array} {l}
   \Psi_L= \left( \begin{array} {l} \nu_e \\ e \end{array} \right)_L ,
     \left( \begin{array} {l} \nu_{\mu} \\ \mu \end{array} \right)_L ,
     \left( \begin{array} {l} \nu_{\tau} \\ \tau \end{array} \right)_L \\
    e_R, \mu_R, \tau_R
\end{array}
\label{psileptonF}
\end{equation}
and the quarks
\begin{equation}
\begin{array} {l}
  \left\{ \begin{array} {l} F=1 \\ F=-1  \end{array} \right. \\
  F=0
\end{array}
\quad
\begin{array} {l}
   \Psi_L= \left( \begin{array} {l} u \\ d' \end{array} \right)_L ,
     \left( \begin{array} {l} c \\ s' \end{array} \right)_L ,
     \left( \begin{array} {l} t \\ b' \end{array} \right)_L .\\
    u_R, d_R, c_R, s_R, t_R, b_R
\end{array}
\label{PsiquarkF}
\end{equation}

The covariant derivative can be rewritten as

\begin{equation}
D_\mu=\partial_\mu+\rm i \frac{g_W}{2}\mathbf{W}_\mu \cdot \tau + \rm i \frac{g_B}{2}F_0 B_{\mu}
\label{DmuFF0}
\end{equation}
where Pauli $2 \times 2$ matrix $\tau$ directly becomes $F$ operator, and $F_0$ substitutes $Y_w$. After mixing $SU(2)$ and $U(1)$, it can be deduced that the charged current part $\mathcal{L}_W$ and the electromagnetic part $\mathcal{L}_{em}$ are same as Equation \ref{LwI3Y} and \ref{LemI3Y} respectively, but the neutral current part becomes
\begin{equation}
   \mathcal{L}_Z=g_Z\bar{\Psi}\gamma^\mu Q_Z\Psi Z_\mu
\label{LZQw}
\end{equation}
where $Q_Z$ is the neutral current weak charge
\begin{equation}
   Q_Z=\frac{c_W^2}{2}F-\frac{s_W^2}{2}F_0 .
\label{QwFF0}
\end{equation}
Interaction vertices of the charged current part and the electromagnetic part are same as Equation \ref{gff} and \ref{Wff} respectively, but one of the neutral current part becomes
\begin{equation}
\begin{array} {l}
    v_f=(\frac{1}{2}-s_W^2)F_f-s_W^2F_{0f} , \qquad  a_f=\frac{1}{2}F_f .
\end{array}
\label{vfafFF0}
\end{equation}

Combining Equation \ref{eqnQF} and \ref{QwFF0} we can obtain
\begin{equation}
\begin{array} {l}
   \left( \begin{array} {l} Q \\ Q_Z \end{array} \right) =
   \frac{1}{2}
   \left( \begin{array} {c} 1 \\ c_W^2 \end{array} \right.
   \left. \begin{array} {c} 1 \\ -s_W^2 \end{array} \right)
   \left( \begin{array} {l} F \\ F_0 \end{array} \right) .
\end{array}
\label{FF0QwQ}
\end{equation}
In consequence, the charge ($Q$, $Q_Z$) of electromagnetic interaction and neutral current weak interaction, which can be combined as neutral current electroweak interaction, can be deduced from mixture of $F$ and $F_0$ via $\theta_W$, and $F_0$ works as the unit electroweak charge. Conservation of the charge ($Q$, $Q_Z$) is originated from conservation of $F$ and $F_0$.

Furthermore, we can take following vectors and tensor which combine constants and parameters of neutral current electroweak interaction\\
electroweak charge vector
\begin{equation}
\begin{array} {l}
   \mathbf{Q_{eZ}}=
   \left( \begin{array} {l} ~~eQ \\ g_ZQ_Z \end{array} \right)
\end{array}
\label{QeZ}
\end{equation}
electroweak fermion quantum number vector
\begin{equation}
\begin{array} {l}
   \mathbf{F_{eZ}}=
   \left( \begin{array} {l} F \\ F_0 \end{array} \right)
\end{array}
\label{FeZ}
\end{equation}
and transformation tensor
\begin{equation}
\begin{array} {l}
   \mathbf{C} =
   \left( \begin{array} {c} 1 \\ a_w \end{array} \right.
   \left. \begin{array} {c} 1 \\ -t_w \end{array} \right)
\end{array}
\label{C}
\end{equation}
where $t_W=\rm tan\theta_W$ and $a_W=\rm atan\theta_W$, and the relation between $\mathbf{Q_{eZ}}$ and $\mathbf{F_{eZ}}$ is gained
\begin{equation}
\begin{array} {l}
   \mathbf{Q_{eZ}} =
   \frac{1}{2}e
   \mathbf{C}\mathbf{F_{eZ}} .
\end{array}
\label{QeZCF}
\end{equation}

Finally, by combining electromagnetic field and  neutral current weak field into neutral current electroweak field

\begin{equation}
\begin{array} {l}
   \mathbf{X}=
   \left( \begin{array} {l} A_\mu \\ Z_\mu \end{array} \right) .
\end{array}
\label{Xmu}
\end{equation}
Lagrangian of electroweak interaction can be rewritten, including two parts
\begin{equation}
-\mathcal{L}_{Wff}=\mathcal{L}_W+\mathcal{L}_{eZ}
\label{LwffeZ}
\end{equation}
where part of neutral current electroweak interaction is
\begin{equation}
\mathcal{L}_{eZ}=\mathcal{L}_{em}+\mathcal{L}_{Z}=\bar{\Psi}\gamma^\mu Q_{eZ}^\nu\Psi X_{\mu\nu}
\label{LeZ}
\end{equation}
where $\nu$=1, 2.

\section{Conclusion and Discussion}

It was regarded historically that two sets of the traditional additive quantum numbers are independent: the traditional flavor-related additive quantum numbers were for strong interaction, and the additive quantum numbers in the electroweak theory were for electroweak interaction. Practically, according to the results in this paper, two sets of the traditional additive quantum numbers are not independent, but equivalent by introducing $F$ and $F_0$. Both $F$ and $F_0$ are conserved in electromagnetic interaction, weak interaction and strong interaction. The new nomenclature of the two quantum numbers in the electroweak theory, i.e., the fermion quantum number ($F$) and the unit electroweak charge ($F_0$), more appropriately reflects the essence of the additive quantum numbers than the old one, i.e., the weak isospin ($I_{w3}$) and the weak hypercharge ($Y_w$). We can unify the traditional flavor-related additive quantum numbers and the additive quantum numbers in the electroweak theory by using only two numbers: $F$ and $F_0$. The new concept system, only including $F$ and $F_0$, is simpler and clearer in the physical connotation than the old concept system.

One conservation law is intimately connected with one certain symmetry which is originated from one indistinguishability before and after a certain transformation. For example, SO(3) group describes weak interaction and conservation of electric charge (present progress in SO(3) can be seen, e.g., in \cite{NE}). Indistinguishability of the fermions results in $F$ conservation, and indistinguishability of relative phases of wave functions of the fermions results in $F_0$ conservation. It can be analogical respectively to the fact that indistinguishability of the protons and neutrons results in $I_3$ conservation in the nuclei, and indistinguishability of relative phases of wave functions of the charged particles results in $Q$ conservation in electromagnetic interaction.

More profoundly, by combining electromagnetic interaction and neutral current weak interaction into neutral current electroweak interaction, the electroweak charge $\mathbf{Q_{eZ}}$ can be deduced from the electroweak fermion quantum number $\mathbf{F_{eZ}}$ via $\theta_W$. Then electroweak interaction is separated into two parts: one is neutral current electroweak interaction which has no change of particle types with only exchange of the neutral bosons, determined by conservation of $F$ and $F_0$, and the other is charged current weak interaction which has change of particle types, i.e., decays and their equivalent reactions, determined by conservation of $F$.

In fact, generality among electromagnetic interaction, neutral current weak interaction and strong interaction is that the flavors of the fermions are not changed (including fermion pair production), and the interactions happen between the fermions via exchanging the neutral bosons (photon, $Z^0$ and gluon respectively). On the the other hand, in charged current weak interaction, including all types of decays and their equivalent reactions, the flavors of the fermions are changed, which means charged current weak interaction happens inside the fermions via emitting or absorbing the charged bosons ($W^\pm$). It is interesting that in neutral current electroweak interaction, the electroweak charge $\mathbf{Q_{eZ}}$ (including $Q$ and $Q_Z$) is determined by the electroweak fermion quantum number $\mathbf{F_{eZ}}$ (including $F$ and $F_0$), meanwhile in charged current weak interaction, the selection rules of the decays and their equivalent reactions are determined by conservation of $F$~\cite{Ma2}. It means that $F$ builds a bridge between neutral current electroweak interaction and charged current weak interaction. In the unified theory of electromagnetic interaction, weak interaction and strong interaction, maybe we should separate  different interactions into the flavor-unchanged part and the flavor-changed part, and $F$ should perform a function as a chain between two parts.

\section*{Acknowledgements}
 This research was funded by National Natural Science Foundation of China (NSFC, No. 12320101005).

\end{document}